\documentclass[%
 aip,
 amsmath,amssymb,
 reprint,%
]{revtex4-1}
\usepackage{xcolor}
\usepackage{graphicx}
\usepackage{dcolumn}
\usepackage{bm}
\usepackage{upgreek} 
\usepackage[utf8]{inputenc}
\usepackage[T1]{fontenc}
\usepackage{mathptmx}
\usepackage{etoolbox}

\makeatletter
\def\@email#1#2{%
 \endgroup
 \patchcmd{\titleblock@produce}
  {\frontmatter@RRAPformat}
  {\frontmatter@RRAPformat{\produce@RRAP{*#1\href{mailto:#2}{#2}}}\frontmatter@RRAPformat}
  {}{}
}%
\makeatother
\begin{document}

\preprint{AIP/123-QED}

\title{Magnons in $\alpha$-Fe$_2$O$_3$ microdisks: from thermal to coherent excitation}
\author{Jinlong Wang}
\thanks{These authors contributed equally to this work.}
\affiliation{%
International Quantum Academy, Shenzhen 518048, China
}%

\author{Katharina Müller}
\thanks{These authors contributed equally to this work.}
\affiliation{%
Walther-Mei$\beta$ner-Institut, Bayerische Akademie der Wissenschaften, 85748 Garching, Germany
}%
\affiliation{%
Physics Department, TUM School of Natural Sciences, Technische Universität München, 85747 Garching, Germany}

\author{Haishen Peng}
\thanks{These authors contributed equally to this work.}
\affiliation{%
International Quantum Academy, Shenzhen 518048, China
}%

\author{Janine Gückelhorn}
\affiliation{%
Walther-Mei$\beta$ner-Institut, Bayerische Akademie der Wissenschaften, 85748 Garching, Germany
}%
\affiliation{%
Physics Department, TUM School of Natural Sciences, Technische Universität München, 85747 Garching, Germany}

\author{Wenjie Song}
\affiliation{%
International Quantum Academy, Shenzhen 518048, China
}%

\author{Lutong Sheng}
\affiliation{%
International Quantum Academy, Shenzhen 518048, China
}%

\author{Jilei Chen}
\affiliation{%
International Quantum Academy, Shenzhen 518048, China
}%

\author{Song Liu}
\affiliation{%
International Quantum Academy, Shenzhen 518048, China
}%

\author{Dapeng Yu}
\affiliation{%
International Quantum Academy, Shenzhen 518048, China
}%

\author{Stephan Geprägs}
\email{stephan.gepraegs@wmi.badw.de}
\affiliation{%
Walther-Mei$\beta$ner-Institut, Bayerische Akademie der Wissenschaften, 85748 Garching, Germany
}%

\author{Matthias Althammer}
\email{matthias.althammer@wmi.badw.de}
\affiliation{%
Walther-Mei$\beta$ner-Institut, Bayerische Akademie der Wissenschaften, 85748 Garching, Germany
}%
\affiliation{%
Physics Department, TUM School of Natural Sciences, Technische Universität München, 85747 Garching, Germany}

\author{Haiming Yu}
\email{haiming.yu@buaa.edu.cn}
\affiliation{%
Fert Beijing Institute, MIIT Key Laboratory of Spintronics, School of Integrated Circuit Science and Engineering, Beihang University, Beijing 100191, China
}%
\affiliation{%
International Quantum Academy, Shenzhen 518048, China
}%

\date{\today}

\begin{abstract} 

Hematite is a prototypical antiferromagnetic insulator for spintronic and magnonic applications, yet controlled studies of coherent magnons in thin films remain challenging due to strain sensitivity and spatial inhomogeneity. Here, we investigate magnon excitations in lithographically patterned hematite microdisks, and obtain reproducible magnon spectra from multiple individual microdisks. Brillouin light scattering reveals bulk-like field-dependent thermal magnon resonances with an ultralow effective damping parameter on the order of $\sim 10^{-5}$. Coherent magnons are further excited using a stripline antenna adjacent to the microdisk, revealing a power-dependent frequency blue shift accompanied by a nonlinear evolution of the coherent response under strong microwave driving. Field-dependent analysis indicates a power-dependent increase of the effective magnetic anisotropy field, suggesting a tunable nonlinear response of antiferromagnetic magnons in laterally confined microdisks. These results establish patterned hematite thin films as a platform for exploring coherent and nonlinear antiferromagnetic magnon dynamics.
\end{abstract}

\maketitle
\begin{figure}
\includegraphics[width=85mm]{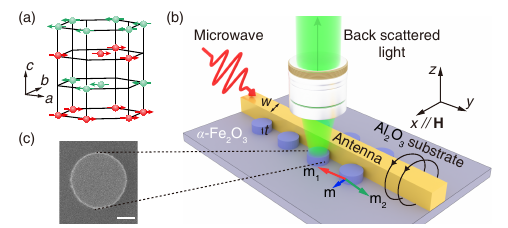}
\caption{Experimental configuration. (a) Crystalline and magnetic structure of $\alpha$-Fe$_2$O$_3$ in the easy-plane phase (oxygen atoms omitted). Arrows represent the orientation of Fe$^{3+}$ spins. (b) Schematic of the experimental setup. A 40-nm-thick (0001)-oriented $\alpha$-Fe$_2$O$_3$ thin film is epitaxially grown on a Al$_2$O$_3$ substrate and patterned into microdisks with a diameter of 3$\,\upmu$m. A microwave stripline antenna (width: 3$\,\upmu$m) is positioned adjacent to the microdisk to coherently drive magnons. An external magnetic field $\textbf{H}$ is applied along the $x$-axis. Magnons are detected using inelastic BLS in a back-scattering geometry, where the incident laser is focused into a Gaussian spot normal to the sample surface (along the $z$-axis). (c) SEM image of a representative $\alpha$-Fe$_2$O$_3$ microdisk (scale bar: 1$\,\upmu$m).}
\label{fig1}
\end{figure}

Antiferromagnetic (AFM) magnons have attracted increasing interest as information carriers owing to their THz-scale dynamics\cite{vaidya2020subterahertz-1,li2020spin-2,zhuang2022excitation-3,yang2024spin-4,leenders2024canted-5}, long-distance transmission\cite{lebrun2018tunable-6,lebrun2020long-7}, and intrinsic robustness against external magnetic perturbations\cite{jungwirth2016antiferromagnetic-8,baltz2018antiferromagnetic-9,han2023coherent-10,Jiahao}. Among candidate materials, hematite ($\alpha$-Fe$_2$O$_3$) is a paradigmatic AFM insulator that has emerged as a promising platform for magnonics\cite{lebrun2018tunable-6,lebrun2020long-7,morrish1994canted-11,pincus1960theory-12,wang2021spin-14,boventer2021room-15,wang2023long-17,hamdi2023spin-18,el2023antiferromagnetic-19,sheng2025control-20,el2025probing-21,chen2025observation-22}. It crystallizes in a corundum structure. Above the spin reorientation (Morin) transition temperature ($T_M$), the Fe$^{3+}$ magnetic moments lie within the (0001) basal plane and are coupled antiferromagnetically along the $c$-axis [Fig.~\ref{fig1}(a)]\cite{morin1951electrical-crystal,besser1967magnetocrystalline-crystal}. This easy-plane configuration is stabilized by magnetocrystalline anisotropy favoring in-plane alignment. Furthermore, the two AFM spin sublattices are slightly canted due to the Dzyaloshinskii–Moriya (DM) interaction\cite{moriya1960anisotropic}, resulting in a spontaneous weak ferromagnetic moment that enables efficient coupling to external microwave and optical probes\cite{wang2021spin-14,boventer2021room-15,wang2023long-17,hamdi2023spin-18}. This enables the study of spin dynamics mainly carried out in bulk hematite demonstrating exceptional dynamical properties, including ultralow magnetic damping\cite{hamdi2023spin-18}, high magnon velocities and long-range spin transport\cite{wang2023long-17}, facilitating the demonstration of fundamental wave phenomena such as magnon nonreciprocity\cite{el2023antiferromagnetic-19} and interference\cite{sheng2025control-20}.

\begin{figure*}[t]
\includegraphics[width=170mm]{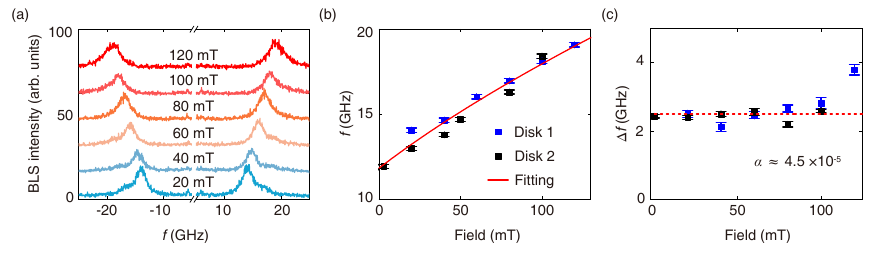}
\caption{Thermal magnon characterization and damping extraction. (a) Thermal magnon spectra of a representative $\alpha$-Fe$_2$O$_3$ microdisk (disk 1) measured at various in-plane magnetic fields $H_\text{ext}$. (b) Resonance frequencies extracted from the peak positions for two microdisks (disk 1 and disk 2) as a function of magnetic field. The solid line is fit to the low-frequency magnon mode of hematite [Eq.~(\ref{FMR})], from which the effective anisotropy field $\upmu_0 H_\text{a}^\text{eff} \approx 90 \pm 5\,\upmu$T, is determined. (c) Field dependence of the frequency-domain linewidth extracted from the thermal BLS spectra. The dashed horizontal line indicates the average linewidth; the larger scatter at high field is discussed in the SM.}
\label{fig2}
\end{figure*}

Significant efforts have been devoted to translating these bulk-like properties into thin-film architectures\cite{wimmer2020observation-13,jani2021antiferromagnetic-16,cheng2019anisotropic-SMR,cheng2020electrical-SOT,han2020birefringence,hattori2024inter,fritjofson2025probing}, which are essential for scalable device integration. In $\alpha$-Fe$_2$O$_3$ thin films, the incoherent generation and detection of magnons have been extensively explored via spin-orbit torque switching\cite{cheng2020electrical-SOT,zhang2022control-SOT2,cogulu2022quantifying-sot3}, spin Hall magnetoresistance\cite{cheng2019anisotropic-SMR,fischer2020large-SMR2,cheng2023unidirectional-SMR3}, and nonlocal transport geometries\cite{wimmer2020observation-13,guckelhorn2023observation-hanle,ross2019propagation-nanolett}, enabling the observation of phenomena such as the magnon Hanle effect\cite{wimmer2020observation-13,guckelhorn2023observation-hanle} and magnetic birefringence\cite{han2020birefringence}. However, achieving stable coherent magnon excitations in thin films remains a formidable challenge\cite{hamdi2023broadband}. The magnetic response of hematite films is highly sensitive to growth-induced strain, which can drastically modify the magnetic anisotropy or even suppress the Morin transition\cite{scheufele2023impact-althammer,jani2021reversible-23,liu2025anisotropic-24}. These factors, combined with the presence of micrometer-scale magnetic domains\cite{ross2019propagation-nanolett}, introduce significant spatial inhomogeneity. Consequently, coherent excitations in thin films are often poorly resolved at low magnetic fields, typically necessitating large bias fields to reach a nearly uniform magnetic state at frequencies spanning hundreds of GHz\cite{hattori2024inter,fritjofson2025probing}. As a result, the coherent dynamics of hematite thin films in the low-field regime remain largely unexplored.

In this work, we demonstrate reproducible magnon spectroscopy in hematite thin films by patterning them into micrometer-scale disks. \textcolor{black}{The lateral confinement reduces spatial averaging over large-scale magnetic inhomogeneities present in the extended film, enabling the magnetic response of individual microdisks to be resolved more clearly.} We use micro-focused Brillouin light scattering (BLS) to investigate the thermal and coherent magnon spectra of individual disks. We analyze the magnetic field-dependence of resonance frequency and linewidth and compare these results to bulk values. For coherently excited magnons, we also look into the effects of microwave power. Our results establish patterned hematite thin films as a controllable platform for investigating nonlinear AFM dynamics, bridging the gap between fundamental bulk physics and scalable thin-film implementations.

Epitaxial 40 nm thick $\alpha$-Fe$_2$O$_3$ thin films are grown on (0001)-oriented single-crystalline sapphire (Al$_2$O$_3$) substrates via pulsed laser deposition [Fig.~\ref{fig1}(b)], with growth parameters detailed in our previous work \cite{scheufele2023impact-althammer}. The high crystalline quality of the films is confirmed by high-resolution X-ray diffraction (see supplementary material (SM)). Vibrating sample magnetometry measurements yielded a magnetization of 4$\pm$0.5 kA/m at a magnetic field of 200 mT at room temperature (see SM), consistent with bulk values\cite{wang2023long-17,chen2025observation-22}. \textcolor{black}{Temperature-dependent magnetization measurements show no signature of a Morin transition over the investigated temperature range, indicating that the film remains in the easy-plane weak-ferromagnetic phase relevant to the BLS measurements (see SM).} To reduce the influence of large-scale magnetic inhomogeneity and enable the characterization of individual magnetic structures, the films are patterned into microdisks with a diameter of 3$\,\upmu$m [Fig.~\ref{fig1}(c)]. \textcolor{black}{The effect of lateral patterning is supported by spatially resolved nitrogen-vacancy stray-field measurements, which reveal reduced magnetic inhomogeneity in the patterned microdisk regions compared with the extended film (see SM).} Magnon characterization is performed using inelastic BLS with a 532 nm single-frequency green laser. The BLS measurements are conducted in a back-scattering geometry with the incident beam normal to the sample plane ($k \approx 0$ rad/$\upmu$m). The laser is focused into a 2$\,\upmu$m diameter spot on the microdisk. An external magnetic field $\textbf{H}$ is applied in-plane along the $x$-axis. \textcolor{black}{Under identical BLS measurement conditions, clear magnon resonances are observed from multiple microdisks, whereas no detectable resonance signal is obtained at multiple different positions on the unpatterned film (see SM), supporting the improved reproducibility associated with lateral patterning.} For coherent magnon excitation, a 3$\,\upmu$m wide Ti(3 nm)/Au(80 nm) stripline antenna is deposited adjacent to the microdisk, with an edge-to-edge separation of $0.5~\upmu$m, and connected to a vector network analyzer (VNA). The microwave current injected into the stripline generates a dynamic magnetic field $h_\text{rf}$, driving the coherent precession of the hematite spins.

\begin{figure*}[t]
\includegraphics[width=170mm]{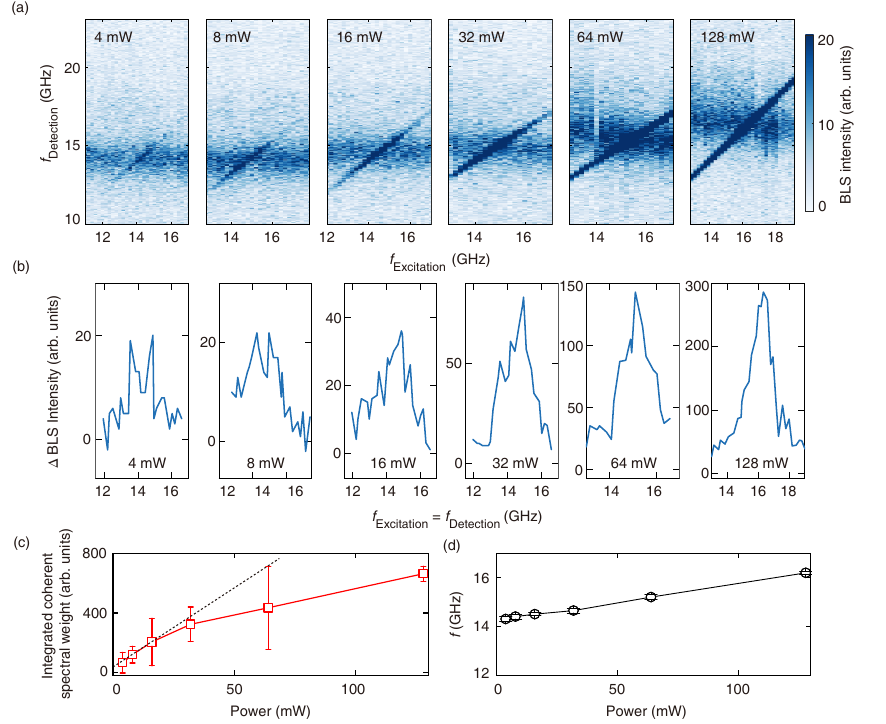}
\caption{Enhancement of BLS signal from coherently excited magnons. (a) BLS intensity map as a function of microwave frequency and power at a fixed in-plane magnetic field of 40 mT. Microwave excitation is applied through the stripline antenna, with the microwave power ranging from 4 to 128 mW. (b) Coherent magnon signal extracted along the resonance trajectory in (a) after subtracting the power-dependent off-resonant background contribution (see SM). (c) Power dependence of the integrated coherent spectral weight extracted from the resonance spectra at each excitation power, showing a transition from an approximately linear response at low power to a nonlinear increase at higher power. (d) Resonance frequency obtained from Lorentzian fits to the spectra in (b) at different excitation powers, showing the power-dependent frequency hardening.}
\label{fig3}
\end{figure*}

Thermal magnon excitations are characterized using BLS at normal incidence. By maintaining the laser beam focus on a selected microdisk and sweeping the in-plane magnetic field $H_\text{ext}$ from 20 to 120 mT, we obtain the field-dependent thermal magnon spectra shown in Fig.~\ref{fig2}(a). Each resonance peak is fitted with a Lorentzian function (see SM) to extract the resonance frequency ($f_\text{res}$) and the full-width at half-maximum linewidth ($\Delta f_\text{res}$). As shown in Fig.~\ref{fig2}(b) and (c), similar field-dependent resonance behavior is observed across multiple microdisks. In the easy-plane phase of $\alpha$-Fe$_2$O$_3$, the low-frequency branch provides the dominant contribution to the observed BLS magnon spectrum. In this mode, the canted moments precess elliptically around their equilibrium positions. The resonance frequency (for $k \approx 0$ rad/$\upmu$m) is described by the relation\cite{elliston1968some-afmr,velikov1969antiferromagnetic-afmr}:
\begin{equation}\label{FMR}
f = \frac{\gamma \upmu_0}{2\pi} \sqrt{H(H + H_\text{D}) + 2H_\text{E} H_\text{a}^\text{eff}}
\end{equation}
where $\gamma$ is the gyromagnetic ratio, $\upmu_0$ is the vacuum permeability, and $H_\text{E}$, $H_\text{D}$, and $H_\text{a}^\text{eff}$ represent the exchange, Dzyaloshinskii-Moriya, and effective in-plane anisotropy fields, respectively. By adopting the values for $\upmu_0H_\text{E} = 1000$ T and $\upmu_0H_\text{D} = 2.22$ T \cite{boventer2021room-15,searle1968magnetic-parameter1,fink1964resonance-parameter2}, we extract by a fit to our data in Fig. 2(b) an effective in-plane anisotropy of $ \upmu_0H_\text{a}^\text{eff} \approx 90 \pm 5\,\upmu$T, which is in agreement with bulk hematite values\cite{wang2021spin-14,boventer2021room-15,wang2023long-17}. \textcolor{black}{To characterize magnetic dissipation, we analyze the frequency-domain linewidth $\Delta f_\text{res}$ extracted from the thermal BLS spectra. For the low-frequency branch, the damping-related linewidth is approximately field independent\cite{fink1964resonance-parameter2,hamdi2023spin-18},
\begin{equation}\label{damping}
\Delta f_\text{res}\simeq 2{\alpha }_\text{eff}\frac{\gamma }{2\pi }\mu_0H_\text{E} 
\end{equation}
As shown in Fig.~\ref{fig2}(c), the measured linewidth remains of the same order over most of the investigated field range, with an average value of approximately 2.5 GHz, corresponding to $\alpha_\text{eff} \approx 4.5 \times 10^{-5}$. This low effective damping parameter is comparable to values reported for high-quality bulk crystals\cite{hamdi2023spin-18}. We note that the measured BLS linewidth also contains unresolved extrinsic broadening contributions such as finite wave-vector acceptance, instrumental resolution, and residual magnetic inhomogeneities. Therefore, the extracted value should be regarded as an upper estimate of the effective damping parameter. A detailed discussion of these linewidth contributions is provided in the SM, Sec. D.}

To further explore driven dynamics beyond thermal excitations, we investigate coherently excited magnons using the stripline antenna described in Fig.~\ref{fig1}(b). The antenna is connected to a VNA, providing microwave excitation with power ranging from 4 to 128 mW. The external magnetic field is fixed at 40 mT. Figure~\ref{fig3}(a) shows the BLS intensity as a function of microwave frequency and excitation power. At low power (4 mW), the coherent-driven signal is comparable to the thermal background, as indicated by a weak diagonal feature within the resonance frequency range. As the microwave power increases, a pronounced diagonal signal develops, indicating efficient coherent magnon excitation. Figure~\ref{fig3}(b) highlights this enhancement after subtraction of the power-dependent off-resonant background (see SM). \textcolor{black}{To quantify the evolution of the coherent response with microwave power, we analyze the resonance spectra at each excitation power and extract the integrated coherent spectral weight. As shown in Fig.~\ref{fig3}(c), the coherent spectral weight exhibits an approximately linear dependence at low microwave powers and gradually deviates from this behavior at higher powers, demonstrating a nonlinear enhancement of the coherent magnon response. This nonlinear behavior reflects the modification of the driven magnon dynamics under strong microwave excitation. The characteristic power scale of this nonlinear response may depend on the microdisk geometry, since lateral confinement and boundary conditions modify the discrete magnon spectrum and spatial mode profiles of confined modes\cite{haghshenasfard2017suhl,english2003modulational}.}

\begin{figure}
\includegraphics[width=85mm]{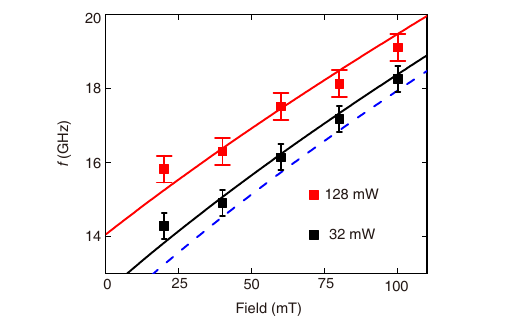}
\caption{Magnetic field dependence of the resonance frequency under coherent excitation at microwave powers of 32 mW (black squares) and 128 mW (red squares). Solid lines represent fits to the low-frequency mode using effective anisotropy fields $H_\text{a}^\text{eff}$ = 100$~\pm~2~\upmu$T (black) and 125$~\pm~3~\upmu$T (red), respectively. The blue dotted line shows the reference fit with $H_\text{a}^\text{eff}$ = 90$~\pm~5~\upmu$T, corresponding to the value extracted from thermal magnon measurements.}
\label{fig4}
\end{figure}

In addition to the nonlinear intensity response, Lorentzian fits to the spectra yield the resonance frequency as a function of excitation power [Fig.~\ref{fig3}(d)]. The resonance frequency systematically increases with microwave power, indicating a pronounced frequency blue shift. This trend is further confirmed by the field-dependent measurements in Fig.~\ref{fig4}, with the corresponding raw-data colormaps provided in Sec. G of the SM. By fitting the field-dependent dispersion, we extract $H_\text{a}^\text{eff}$. The thermal magnon measurements yield $H_\text{a}^\text{eff}=90\pm5~\upmu$T, which increases to $100\pm2~\upmu$T at 32 mW and $125\pm3~\upmu$T at 128 mW. Thus, the observed hardening can be phenomenologically described by a power-dependent increase in $H_\text{a}^\text{eff}$, indicating a modification of the effective magnetic energy landscape under coherent microwave driving. \textcolor{black}{To assess whether the observed hardening can arise from microwave-induced heating, finite-element simulations are performed and indicate a local temperature rise of only $\sim5$ K at the highest microwave power (see SM). Temperature-dependent BLS measurements from 295 to 320 K demonstrate that the low-frequency resonance remains nearly unchanged and, if anything, exhibits a slight softening with increasing temperature, consistent with previous antiferromagnetic resonance measurements\cite{velikov1969antiferromagnetic-afmr}. This behavior is opposite to the pronounced microwave-power-induced hardening. Since the temperature-dependent measurements inherently include thermal-expansion and associated magnetoelastic effects of the film/substrate structure, these results indicate that microwave-induced heating is unlikely to be the dominant origin of the observed blue shift.} Interestingly, the observed hardening contrasts with the frequency softening reported for large-amplitude dynamics in bulk antiferromagnets\cite{el2025probing-21}. In the confined microdisks, modified boundary conditions and the discrete mode structure may influence the driven magnon response. Irrespective of its precise microscopic origin, the pronounced power-dependent hardening reveals a nonlinear response of the microdisk resonance under coherent driving. A complete microscopic description of this nonlinear hardening remains an interesting question for future work.

In summary, we demonstrate controllable thermal and coherent magnon excitation in patterned hematite microdisks using BLS. The thermal magnon response shows bulk-like behavior with an ultralow effective damping parameter on the order of $\sim 10^{-5}$. Under coherent excitation, we observe a pronounced frequency blue shift together with a nonlinear increase of the coherent magnon response. Field-dependent measurements reveal a power-dependent increase in the effective in-plane anisotropy field, indicating tunable nonlinear dynamics in confined antiferromagnetic structures. Our results establish patterned $\alpha$-Fe$_2$O$_3$ thin films as a platform where the favorable magnonic properties known from bulk hematite can be preserved and actively controlled, opening a pathway toward scalable AFM magnonic devices.

See supplementary material for the structural and magnetic characterization of the $\alpha$-Fe$_{2}$O$_{3}$ thin film, effect of lateral patterning on magnetic homogeneity and magnon response, Lorentzian fit of the thermal magnon resonance, contributions to the measured BLS linewidth, extraction of the coherent magnon contribution from the BLS spectra, microwave-induced heating and thermal strain, and microwave-driven resonance spectra as a function of magnetic field.

\begin{acknowledgments}
We wish to acknowledge the support by the National Key Research and Development Program of China, Grants No. 2022YFA1402801; NSF China under Grants Nos. 12604179, 12525406, 12474104, 12504144; MA and KM acknowledge support via the European Research Council (ERC) under the European Union’s Horizon Europe research and innovation programme (Consolidator Grant ”POSA”, Grant Agreement No. 101171325) and the  Deutsche Forschungsgemeinschaft (DFG, German Research Foundation) via Germany’s Excellence Strategy EXC-2111-39081486.
\end{acknowledgments}


\section*{Data Availability Statement}
The data that support the findings of this study are available from the corresponding author upon reasonable request.

\nocite{*}
\bibliography{hematiteThinFilmBiB}
\end{document}